\documentclass[dvips]{article}
\usepackage{icrctc07}

\title{Observations of Pulsar Wind Nebulae with the VERITAS Array 
of Imaging Atmospheric Cherenkov Telescopes}
\shorttitle{Observations of PWNe with VERITAS}
\authors{A. Konopelko$^{1}$, for the VERITAS collaboration$^2$}
\shortauthors{A. Konopelko and et al}
\afiliations{$^1$Purdue University, Department of Physics, 525 Northwestern Avenue, 
West Lafayette, IN 47907-2036, $^2$For full author list see G. Maier, "VERITAS: Status 
and Latest Results", these proceedings}
\email{akonopel@purdue.edu}

\abstract{Many of the recently discovered galactic very high-energy (VHE) $\gamma$-ray sources 
are associated with Pulsar Wind Nebulae, which is the most populous Galactic source category at 
TeV energies. The extended synchrotron nebulae of these objects observed in the X-ray band 
are a hallmark of the relativistic winds, generated by the young, energetic neutron stars, that interact 
with the matter ejected by the supernova explosion and the surrounding interstellar gas. Relativistic 
electrons, or protons, accelerated in the pulsar winds, or at their shock boundaries, interact with the 
magnetic field and low energy seed photons to produce the observed VHE 
$\gamma$-ray emission. The VERITAS array of four imaging atmospheric Cherenkov telescopes 
was designed to study astrophysical sources of $\gamma$ rays in the energy domain from about 
100 GeV up to several tens of TeV. The sensitivity of the VERITAS array allows detailed studies of 
the morphology and spectral features of $\gamma$-ray emission from PWNe. Three northern 
sky PWNe, G75.2+0.1, G106.6+2.9, and 3C58, were observed with VERITAS 
during 2006. No evidence for TeV $\gamma$-ray emission at the position of the pulsar 
associated with these PWNe is demonstrated. 
}

\begin{document}
\maketitle

%Begin the section.

\section{Introduction}
\vspace{-2.5mm}

Charged particles accelerated in the vicinity of a rapidly rotating neutron star, or pulsar,
flow out into the interstellar medium and encounter the supernova ejecta from the pulsar's birth
event and form a shock. The shock may further 
enhance the acceleration of the particles which can then attain relativistic speeds. 
This interaction between the accelerated charged particles and the 
surrounding medium produces a pulsar wind nebula (PWN). PWNe are often 
observable at wavelengths from the radio through the $\gamma$-ray. Around the youngest, most 
energetic pulsars, the radio emitting regions of these nebulae are rather 
amorphous, whereas the X-ray emitting regions can be highly structured 
\cite{ref1}. The high spatial resolution of the {\it Chandra} X-ray 
Observatory has made it possible to resolve the structures of PWNe.
The presence of a PWN can also be inferred spectrally. For instance, a 
non-thermal component is often seen in the ASCA and INTEGRAL 
observations of pulsars, such as PSR~B1509-58 and PSR~B1046-58. 
This seems to suggest that PWNe are a common phenomenon for all 
energetic pulsars \cite{ref2}.

\begin{table*}[t]
\centering
\begin{tabular}{l|cccccccc}\hline
Pulsar & PWN & P (ms) & B ($10^{12}$G) & t (kyr) & 
$\rm \dot{E}$ ($10^{36}$erg/s) & d (kpc) & $\rm F_\gamma$($10^{-11}$erg/s)  \\ \hline
J2021+3651 & G75.2+0.1   & 104 & 3.2 & 17 & 3.4 & 12 & 7 \\
J2229+6114 & G106.6+2.9 & 51.6 & 2 & 10.5 & 22 & 3 & 4 \\
J0205+6449 & 3C58  & 65 & 3.6 & 5 & 27 & 3.2 & 0.8 \\ \hline
\end{tabular}
\caption{Physical parameters of observed pulsars and their PWNe. }
\label{t1}
\vspace{-4mm}
\end{table*}

It was widely believed that PWNe are potential sources of  
VHE $\gamma$-ray emission. The emission probably arises from inverse Compton (IC) 
scattering of low-energy photons by the relativistic electrons, while the X-ray 
emission is associated with the synchrotron radiation from the same population of electrons. 
The best example of a PWN is the Crab Nebula, which is an established source of 
pulsed $\gamma$-ray emission up to a few GeV detected by EGRET, as well 
as a source of steady TeV $\gamma$ rays observed by a 
number of ground-based Cherenkov detectors and recently with the VERITAS 
array of four imaging atmospheric Cherenkov telescopes \cite{ref3}. The 
TeV emission is thought to originate at the base of its PWN. The H.E.S.S. detector,
located in the southern hemisphere, discovered a number of previously unknown 
$\gamma$-ray sources in the VHE domain above 100 GeV. A total of five of these 
new sources (PSR~B1509-58, Vela~X, ``Kookaburra'', SNR~G0.9+0.1, 
PSR~B1823-13) are apparently associated with PWNe. Such associations 
rest on a positional and morphological match of the VHE $\gamma$-ray source to a 
known PWN at lower energies. It is worth noting that in 
all cases the pulsar is significantly offset from the center of the VHE $\gamma$-ray 
source. This offset could be attributed to the interaction between the PWN and the 
SNR ejecta \cite{ref5}.

A detailed survey of the inner part of the Galactic Plane at VHE
$\gamma$-ray energies has been carried out with H.E.S.S. Fourteen previously unknown, 
extended sources were detected with high significance \cite{ref6}. 
Some of these sources have fairly well-established counterparts at longer wavelengths, 
based exclusively on positional coincidence, but others have none at all. A number 
of models have been proposed regarding the nature of these unidentified VHE $\gamma$-ray 
sources. At present, PWNe and shell-type SNRs are considered the most plausible 
counterparts for the remaining unidentified VHE $\gamma$-ray sources amongst the
numerous possibilities that have been put forward.

\vspace{-2mm}
\section{Targets}
\vspace{-2mm}

Motivated by the growing catalog of TeV PWNe, VERITAS, in 2006, observed 
three northern sky PWNe, G75.2+0.1, G106.6+2.9, and 3C58, associated with young, 
energetic pulsars (see Table~\ref{t1}).

\noindent
{\bf PSR~J2021+3651}. A {\it Chandra} observation showed this 
pulsar to be embedded in a compact, bright X-ray PWN (PWN G75.2+0.1) with the standard torus and 
jet morphology \cite{ref7}. Its X-ray spectrum is well fit by a power-law model with photon 
index $\Gamma$=1.7 and a corresponding 0.3-10~keV flux of 
1.9$\times 10^{-12}\, \rm erg\, cm^{-2}\, s^{-1}$. This young Vela-like pulsar is coincident 
with the EGRET $\gamma$-ray source GeV~2020+3651. Recently, the Milagro 
$\gamma$-ray observatory detected an extended source or multiple unresolved 
sources of $\gamma$ rays at a median-detected energy of 12~TeV \cite{ref8} coincident with
the same region. 
The radio dispersion measure suggests a distance to PWN G75.2+0.1 $\rm d \geq 10$~kpc, 
but this measurement could have been contaminated by the gas in the Cygnus 
region and the true distance may be in fact substantially closer. Presently PWN G75.2+0.1 
is considered to be one of the best candidates for the Milagro source (MGRO J2021+37) and it is 
likely to be seen in the energy range covered by VERITAS.

\noindent
{\bf PSR~J2229+6114}. The {\it Chandra} X-ray image of PWN G106.6+2.9 shows 
an incomplete elliptical arc and a possible jet, similar to the Vela PWN \cite{ref9}. 
PSR~J2229+6114 is a compelling counterpart of the EGRET source 
3EG~J2227+6122. This young, energetic pulsar is second only to the Crab pulsar in spin-down 
power, and it is  substantially more luminous than the Vela pulsar. Given the relatively 
small distance of 3~kpc this pulsar has a very high rank among all pulsars 
in the discriminant $\dot{E}/d^2$. Part of this flux can be converted into a high flux of VHE 
$\gamma$ rays.

\noindent
{\bf 3C58}. 3C58 is a young Crab-like SNR generally accepted as being the 
remnant of the historical supernova SN 1181. A compact object (nebula) at the center of the 
SNR has been resolved in {\it Chandra} X-ray data \cite{ref10}, and is centered on 
PSR~J0205+6449. Given its very high spin-down power, the pulsar is capable of 
supplying the energy of the X-ray nebula, $L_x = 2.9\times 10^{34}\, \rm ergs\, s^{-1}$, 
and may have substantial VHE $\gamma$-ray emission.   
     
The TeV $\gamma$-ray fluxes expected from the PWNe around both PSR~J2021+3651 and 
PSR J2229+6114 in terms of a hadronic-leptonic model for the high-energy 
processes inside the PWNe \cite{ref11} exceed 10\% of the Crab Nebula 
flux above 200~GeV (see Table~1). This suggests that both PSR J2021+3651 and 
PSR J2229+6114 should be detectable with VERITAS 
after rather short exposures. A somewhat lower $\gamma$-ray flux 
of a few percent of the Crab Nebula was predicted for 3C58 \cite{ref11},
however it is still well above the sensitivity limit of the VERITAS detector for a reasonable exposure. 

\vspace{-3mm}
\section{VERITAS Observations and Analysis}
\vspace{-2mm}

\begin{table*}[!t]
\centering
\begin{tabular}{lcccccccccc}\hline
Pulsar & $\rm N_{tel}$  & $\rm N_{runs}$ & $\rm R$ (Hz) & T (hr) & $\Theta$ ($^\circ$) & On & Off & $\alpha_{\small Li\&Ma}$ & S/N ($\sigma$) & {\small U.L. (Crab)}  \\  \hline
{\small J2021+3651} & 2  & 23 &  100  &  8.4  & 30.5 & 189 & 512 & 0.33 & 1.19 & 4.6\% \\
{\small J2229+6114} & 2  & 32 & 88 & 12  & 31.8 & 151 & 543 & 0.25 & 0.19  & 2.7\%  \\
{\small J0205+6449} & 3  & 17 & 147 & 5.3 & 34.2 & 109 & 406 & 0.25 & 0.32 & 2.4\% \\ \hline
\end{tabular}
\caption{Summary of data.  }
\label{t2}
\vspace{-4mm}
\end{table*}

VERITAS is an array of four imaging Cherenkov telescopes sited in Amado, 
Arizona, and dedicated to the detection of VHE $\gamma$ rays with 
energies above 100 GeV. Each telescope has a tessellated mirror with an area of 
$\simeq 110$~m$^2$  and a camera consisting of 499 photomultiplier tubes. The 
first telescope in the array has been operating since February 2005. First stereo 
observations with two telescopes began in April 2006, and the full array of four 
telescopes has been operational since January 2007. A full VERITAS array has the sensitivity of
7~mCrab (5$\sigma$ detection over 50 hour exposure).
%  is currently the most 
%sensitive ground-based 
%VHE $\gamma$-ray instrument in the northern hemisphere. 
The angular resolution 
of better than $0.14^\circ$ and a 3.5$^\circ$ field of view enable VERITAS to detect
and study a variety of compact galactic $\gamma$-ray sources like PWNe.

The VERITAS observations of three PWNe were made while the system was 
under construction. Observations of PSR~J2021+3651 and PSR~J2229+6114 in 
November 2006 were made with a two telescope system. Later a third 
telescope was added to the system and observations of 3C58 in December 2006 
were made with three telescopes. The data were taken mostly in 20~minute runs 
with a few runs of 28~minutes using the {\it wobble} mode. In this mode, the source 
direction is positioned $\pm 0.3^\circ$ (a $\pm 0.5^\circ$ offset was used for 
later observations) in declination or right ascension relative to the center of the 
camera field of view. The sign of the offset was altered in successive runs  to 
reduce systematic effects. The {\it wobble} mode allows on-source observation 
and simultaneous estimation of the background induced by charge cosmic-ray 
particles. This eliminates the need for off-source observations and consequently 
doubles the amount of available on-source time.    

For final analysis only those runs passing the data quality criteria are used. 
The images are calibrated and then cleaned using a two-threshold 
picture/boundary selection procedure which requires a pixel to have a signal 
greater than 5.0 pedestal variances (PV) and a neighboring pixel to have a signal 
larger than 2.5 PV. The pixels with a signal greater than 2.5 PV are included only 
if they have a neighbor with a signal greater than 5.0 PV. After image cleaning
the shower images are parameterized using a standard second-moment approach. 
The shower geometry is reconstructed using stereoscopic techniques with a typical 
angular resolution of  about 0.14$^\circ$ and an average accuracy of  better than 20 m 
in the determination of the shower core location. To ensure that images are not 
truncated by the camera edge, only images with the center of gravity less than 
1.3$^\circ$ from the center of the camera are used in the reconstruction. In addition, 
at least two images are each required to exceed a minimum total signal of 400 
digital counts of the respective flash analog-to-digital converter to ensure that the 
showers are well reconstructed.

After the shower reconstruction, the cosmic-ray background events are rejected 
using standard cuts on mean scaled width and mean scaled length parameters. The number 
of events passing cuts in a circle of standard angular size around the source 
position gives the number 
of on-source (On) counts. The background is estimated using all events passing cuts in a 
number of non-overlapping circles of the same size. The centers of these circles 
are positioned at the wobble offset from the tracking position. The number of 
background regions may vary depending on the actual wobble offset used in the
observation. The use of a larger background region reduces the relative
statistical error on the background measurement. For a given number of on-source 
and background counts acquired after event selection the significance of the excess is calculated 
following the method of Equation (17) in the Li \& Ma technique \cite{ref12}.
It is worth noting that the data have been analyzed using independent analysis packages 
(see \cite{ref16, ref17} for details on the analyses). All of these analyses yield consistent results.

%\begin{figure}[t]
%\vspace{-2mm}
%\includegraphics [width=0.46\textwidth]{bluemap}
%\caption{Sky map ($\delta$: declination, $\alpha$: right ascension) of excess counts 
%in direction of PSR~J2021+3651.  
%}
%\label{fig1}
%\vspace{-3mm}
%\end{figure}

\vspace{-3mm}
\section{Results and Conclusion} 
\vspace{-1mm}

Table~2 summarizes the results of the VERITAS observations of each of the 
individual sources. These objects have been observed with VERITAS for rather 
limited exposure times. The longest exposure (T) of 12~hrs was for PSR~J2229+6114. 
All observations were made at the median zenith angle ($\Theta$) of $\sim 30^\circ$. 
Parameter $\alpha_{Li\&Ma}$ used in the Li \& Ma technique is also given in Table~2. No 
significant excess suggesting TeV $\gamma$-ray emission is evident for any of the observed 
PWN. The 99\% confidence level flux upper limits \cite{ref13} at the pulsar position expressed 
in the flux of the Crab Nebula are shown in Table~\ref{t2}. 

Present VERITAS upper limits for a sample of northern PWNe contradict  
predictions of high VHE $\gamma$-ray fluxes made in \cite{ref11}, which might possibly 
constrain the choice of magnetic field strength within the nebulae. For PSR~J2021+3651  
a very hard $\gamma$-ray spectrum of hadronic origin with a peak at $\sim$10~TeV, which 
is the median-energy of the MILAGRO detection, and sharp fall off at higher energies as 
suggested in \cite{ref14} would still satisfy the VERITAS upper limit. A similar spectrum 
has been observed from PWN Vela X by H.E.S.S. \cite{ref15}. Note, however, that 
present VERITAS upper limits were derived assuming a point like $\gamma$-ray source. 
Estimating the VERITAS upper limit for an extended source seems to be premature  
given that the exact localization and angular extent of the VHE $\gamma$-ray source detected 
by the MILAGRO $\gamma$-ray observatory are still in the process of final evaluation.    

\vspace{-2mm}
\subsection*{Acknowledgments}
\vspace{-1mm}

VERITAS is supported by grants from the U.S. Department of Energy, the
U.S. National Science Foundation and the Smithsonian Institution, by NSERC in
Canada, by PPARC in the U.K. and by Science Foundation Ireland.

\vspace{-3mm}
%This is the reference to .bib file (Whitout .bib!)
%\bibliography{libros1}
%This in the bibtex style, is ok.
\bibliographystyle{plain}

\end{document}